\documentclass{elsart}

\usepackage{amssymb}

\usepackage{graphicx}
\usepackage{dcolumn}
\usepackage{bm}

\begin{document}

\begin{frontmatter}


\title{Fluorescence photons produced in air by extensive air showers} 

\author{Vitor de Souza\corauthref{cor1}}
\author{, Gustavo Medina-Tanco}
\author{and Jeferson A. Ortiz}
\corauth[cor1]{vitor@astro.iag.usp.br}
\address{
Instituto de Astron\^omia, Geof\'isica e Ci\^encias Atmosf\'ericas \\
Universidade de S\~ao Paulo\\ 
Caixa Postal 9638 - CEP 05508-900, S\~ao Paulo - SP, Brasil }

\begin{abstract}
The air fluorescence technique has long been used to detect extensive air
showers and to reconstruct its geometry and energy. The fluorescence
photon yield of an electron in air is of main importance in the
reconstruction procedure.
Historically, the fluorescence yield used in the
reconstruction of the showers is approximated at all energies by that
of an electron with kinetic 
energy of 80 MeV, because these are the most abundant in a shower.
 In this paper, we calculate the
fluorescence yield 
taking into account the energy spectrum of electrons in 
showers initiated by proton and iron nuclei as a function of
height. We compare our results with previous 
calculations based on mono energetic electrons (80 MeV) 
and a difference in excess of 8\% is
found. Finally, the influence of a more realistic fluorescence yield in the shower
energy reconstruction is also discussed. 
\end{abstract}

\begin{keyword}
cosmic rays \sep extensive air showers \sep air fluorescence yield
\sep energy reconstruction \PACS 96.40-z,96.40.Pq,96.40.De
\end{keyword}
\end{frontmatter}

\section{\label{introduction} Introduction}

Fluorescence telescopes have been successfully used to measure
extensive air shower since the Fly's Eye \cite{bib:flyseye} experiment
proved its  efficiency. Presently, the HiRes \cite{bib:hires}, Auger
\cite{bib:auger-nim}, EUSO \cite{bib:euso-artigo}, OWL
\cite{bib:owl-artigo} and  Telescope Array \cite{bib:telescope:array})
Experiments are using or planning to use the fluorescence technique to
study the cosmic ray with energy above $10^{19}$ eV.

The detection of the shower by a fluorescence detector is performed by
measuring the amount of fluorescence light produced in air by the excitation of the
nitrogen molecules as  particles (essentially electrons and positrons)
travel along the atmosphere. The
number of photons detected by the telescope can be converted to the
number of charged particles in the shower at a given height taking into
account the absorption and scattering of the fluorescence light along  the path
from the emission to the telescope and knowing the fluorescence yield
of the particle.

From the number of particles as a function of height one is able to
reconstruct the energy of the shower following the arguments in
reference \cite{bib:flyseye_calib}. Therefore, the fluorescence yield
has a direct influence in the energy estimation of a shower.

The importance of knowing the fluorescence yield of the electron as a
function of energy is underlined by the number of experiments
presently under operation to measure the yield as can be seen in
references
\cite{bib:fy:campinas,bib:fy:airfly,bib:nagano,bib:nagano2}.

From the measurements of the yield in the experiments cited above to
the use of the fluorescence yield in cosmic rays showers a number of
assumptions are made. Specially, the fluorescence yield of the
particles in a shower are considered to be the yield of electrons with
kinetic energy of 80 MeV.

The contribution of other particles, like muons and hadrons and
particles with energy below the simulation energy cut, to the total
productions of fluorescence light in a shower has been extensively studied in a
recent paper by H. Barbosa et al. \cite{bib:hbarbosa}. However, the
mono energetic approximation has survived widely unchallenged.

The scope of the present work is to evaluate the approximation of the
energy spectrum of electron by single energy (80 MeV) in the
evaluation of the fluorescence yield.  We calculate the fluorescence
yield taking into account  the true energy spectrum of electrons (=
electrons + positrons) in showers initiated by proton and iron nuclei
with energy $10^{18}$, $10^{19}$  and $10^{20}$ eV.   Our results show
that the approximation of mono energetic electrons (80 MeV) leads to
an over estimation of the shower energy by a factor of $\sim$ 8\%.

\section{\label{sec:TeoYield} Fluorescence Yield Measurements and
  Theory}

The fluorescence yield of electrons in air is determined by a
competition of two factors: the probability of excitation of a nitrogen
molecule and the probability of its de-excitation via collisions with
mainly oxygen molecules (quenching).

Based on this assumption Bunner \cite{bib:bunner} suggested that the
fluorescence yield (\emph{FY}) should be proportional to the energy deposit
($dE/dx$) and can be parametrized as:

\begin{equation}
FY_{Bunner}(K,\rho,T) \propto \frac{dE}{dx} \times \frac{\rho}{1+\rho B \sqrt{T}}
\end{equation}

where \emph{K} is the kinetic energy of the particle, $\rho$ is the
air density, \emph{B} is a constant and \emph{T} is temperature. The
$dE/dx$ is the energy loss rate per unit path which is dominated by
ionization losses.

More recently, two measurements have been done to determine the
fluorescence yield as a function of the kinetic energy of the
electron. Kamimoto et al. \cite{bib:kakimoto} measured the
fluorescence yield for electrons with energy  1.4, 300, 650 and 1000 MeV
and confirmed the proportionality of the yield with the energy loss
rate ($dE/dx$) as suggested by Bunner \cite{bib:bunner}. 
Another experiment has been
done by Nagano et al. \cite{bib:nagano} and the same proportionality 
between the yield and the $dE/dx$ was verified despite the fact that this
experiment was able to make measurements at a greater number of
wavelength bandwidths than Kakimoto et al. did.

\begin{figure}[]
\centerline{\includegraphics[width=13cm]{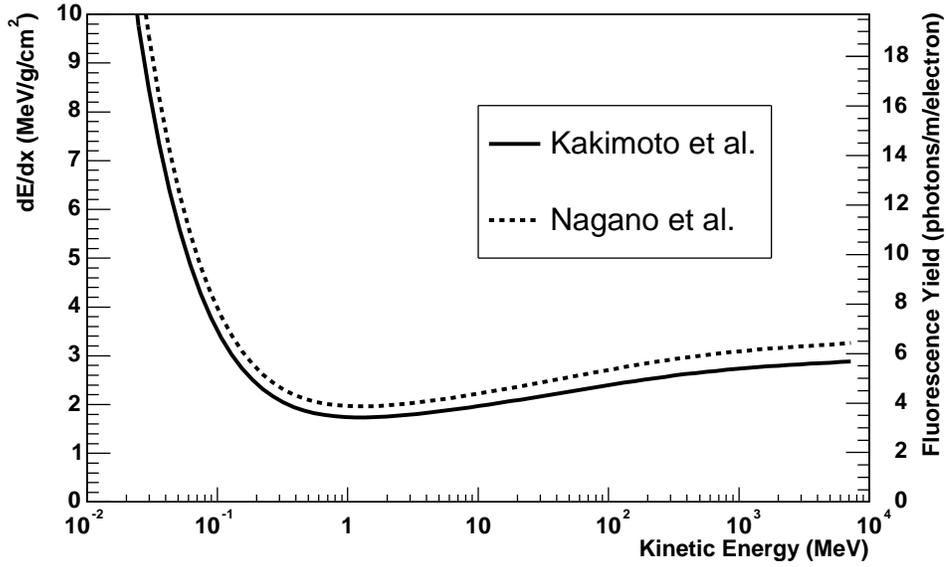}}
\caption{The \emph{dE/dx} of an electron in air as a function of its
  kinetic energy and the corresponding fluorescence yield as suggested
  by Kakimoto et al. and Nagano et al. We have used $T
  = 300$ K and $\rho = 760$  mm  Hg.}
\label{fig:dedx:yield}
\end{figure}

Both measurements suggest that the relation between yield and $dE/dx$
can be given by:

\begin{equation}
FY(K,\rho,T)  = \frac{\left( \frac{dE}{dx} \right)}{ \left( \frac{dE}{dx}
\right)_{K_c}} \times \; \rho \left(\frac{A_1}{1 + \rho B_1 \sqrt{T}}
+ \frac{A_2}{1 + \rho B_2 \sqrt{T}} \right)
\label{eq:yield}
\end{equation}

where $\rho$ is the density in $\mathrm{kg\ m^{-3}}$ and \emph{T} is the
temperature in Kelvin. The constants $A_1$, $A_2$, $B_1$, $B_2$ and
$(dE/dx)_{K_c}$ is the energy loss rate calculated at $K_c$ as given
in Table \ref{tab:const}. At a given pressure 
and temperature Kakimoto et al. and Nagano et al. differ in the
relation between the energy loss rate and the fluorescence yield only
by a constant as can be seen in figure \ref{fig:dedx:yield} where
equation \ref{eq:yield} is plotted for $T = 300$ K and $\rho =
760$  mm  Hg.

\begin{figure}[]
\centerline{\includegraphics[width=13cm]{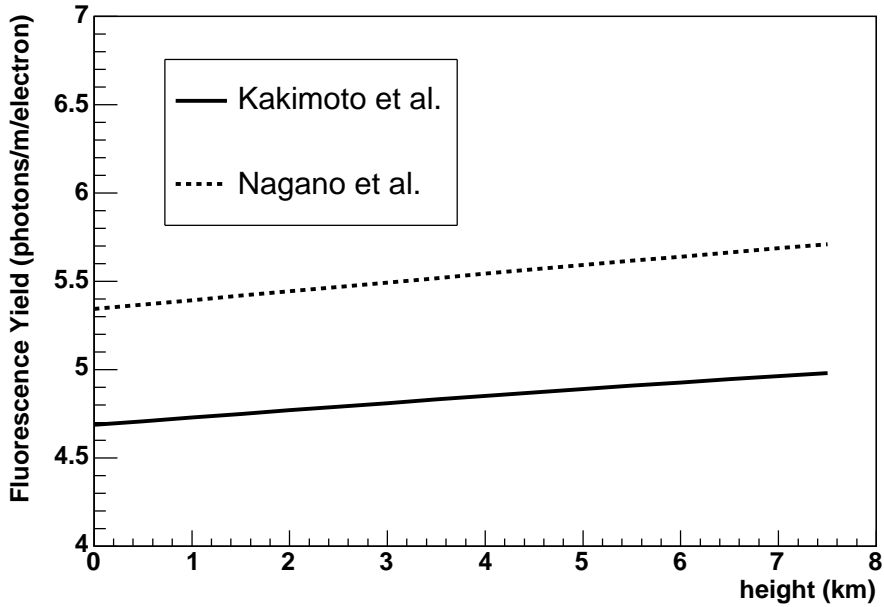}}
\caption{The fluorescence yield as a function of height in the
  atmosphere for electrons with kinetic energy equal to 80
  MeV. Pressure and temperature varies with height according to the
  US Standard Atmospheric Model \cite{bib:bianca:artigo}.}
\label{fig:yield:height}
\end{figure}

Taking into account the variation of the density and temperature of
the atmosphere as a function of height, it is possible to determine at
a given energy how the yield varies with height. Figure
\ref{fig:yield:height} shows the fluorescence yield as a function of
height for electrons with kinetic energy of 80 MeV.

This energy is normally chosen because it is the energy at which the
probability of an electron undergoing pair-production is equal to the
probability of absorption by ionization and therefore the number of
particles in a shower reaches its maximum. However, the energy of the
electrons in a shower has a wide energy distribution. Furthermore, the
energy distribution is different at different heights and for distinct primary
particles.

Nevertheless, the fluorescence yield is a highly dependent function on the kinetic
energy (through $dE/dx$) and, since the energy distribution of the electrons varies with
height, the fluorescence yield must vary with height in a manner different
from that one shown in figure \ref{fig:yield:height}.

Figure \ref{fig:yield:height} shows that, for 80 MeV electrons, the
fluorescence yield is a  slowly varying function of the height in the atmosphere.
The later has been accepted as one of the basic features
on which rests the fluorescence technique. However, as explained above, the
characteristic value of the distribution of yields as a function of
kinetic energy at a given height is not equal to the yield calculated
at the characteristic value of the energy distribution at a given height.

In the next sections we are going to convolute the energy distribution
of electrons for heights varying from the sea level up to 7.5 km and
for each height we are going to calculate the characteristic energy and fluorescence
yield. A distribution of fluorescence yields for each height was
determined allowing us to estimate the confidence level of the
distribution of yields at each height.

\subsection{The influence of different atmospheric profiles in the
  yield}

In applications of the fluorescence yield to measure cosmic rays
showers, a precise knowledge of the fluorescence yield and its
dependences with the kinetic energy of the particle are useless if one
does not have a good measurement of the atmospheric parameters.

This effect has been shown by B. Keilhauer et
al. \cite{bib:bianca:gapnote} and we are going to stress its
importance here by explicitly calculating the fluorescence yield.

B. Keilhauer et al. have measured the atmospheric profile in the
Pierre Auger Observatory site during an entire year. They have used
meteorological radio soundings to determine the temperature and
pressure variation as a function of height. Their results are published
in \cite{bib:bianca:artigo} and we are going to use them in the
following calculation.
 
\begin{figure}[]
\centerline{\includegraphics[width=13cm]{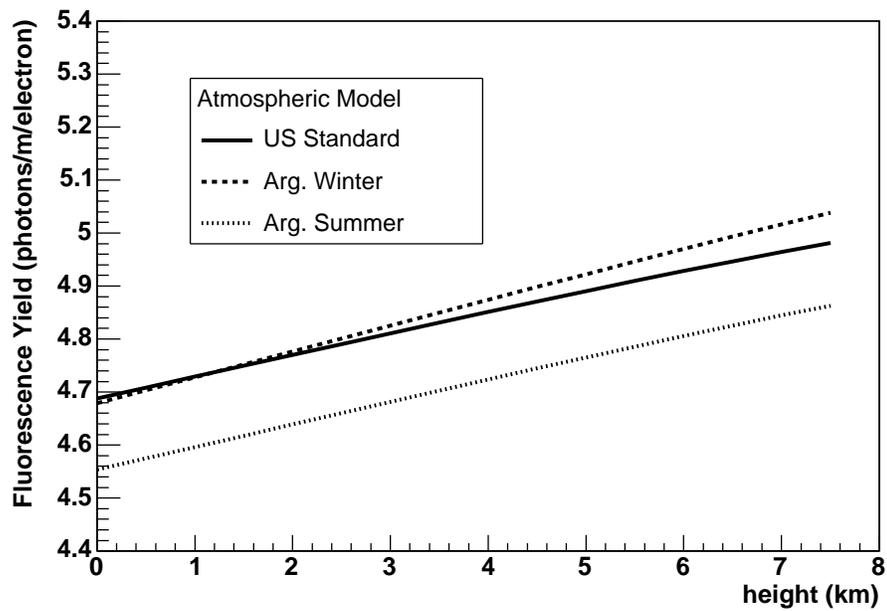}}
\caption{The fluorescence yield as a function of height in the
  atmosphere for electrons with kinetic energy equal to 80
  MeV according to Kakimoto et al. suggestion. Pressure and
  temperature varies with height according to the 
  models published by B. Keilhauer et al. \cite{bib:bianca:artigo}. Arg. Winter and
  Summer are the average measurements for the Pierre Auger Southern site.}
\label{fig:yield:atm}
\end{figure}

Knowing the pressure and temperature variation as a function of height
it is possible to calculate the variation of the fluorescence yield as
a function of height by using equation \ref{eq:yield}. Figure
\ref{fig:yield:atm} shows the value of the fluorescence yield for
three atmospheric models according to Kakimoto et al.. Seasons effects
 can change the fluorescence yield 
by as much as 4 \% \cite{bib:bianca:gapnote}.

\section{The fluorescence yield according to the electron energy spectrum}
\label{sec:analitic}

In order to evaluate the influence of the distribution of the
electron's kinetic energy we have done Monte Carlo simulations with
the program CORSIKA \cite{bib:corsika} linked to the QGSJET
\cite{bib:qgsjet} interaction model. We have  defined  observational levels
varying from sea level up to 7.5 km and determined the energy spectrum
of the electrons for each level.

The simulations were performed with the thinning factor of $10^{-6}$ and
the energy cuts for hadrons, muons, electron and photons were
respectively set to 50 MeV, 50 MeV, 50 keV and 50 keV. Ten
vertical showers were simulated for each configuration of energy and primary
particle.

\begin{figure}[]
\centerline{\includegraphics[width=13cm]{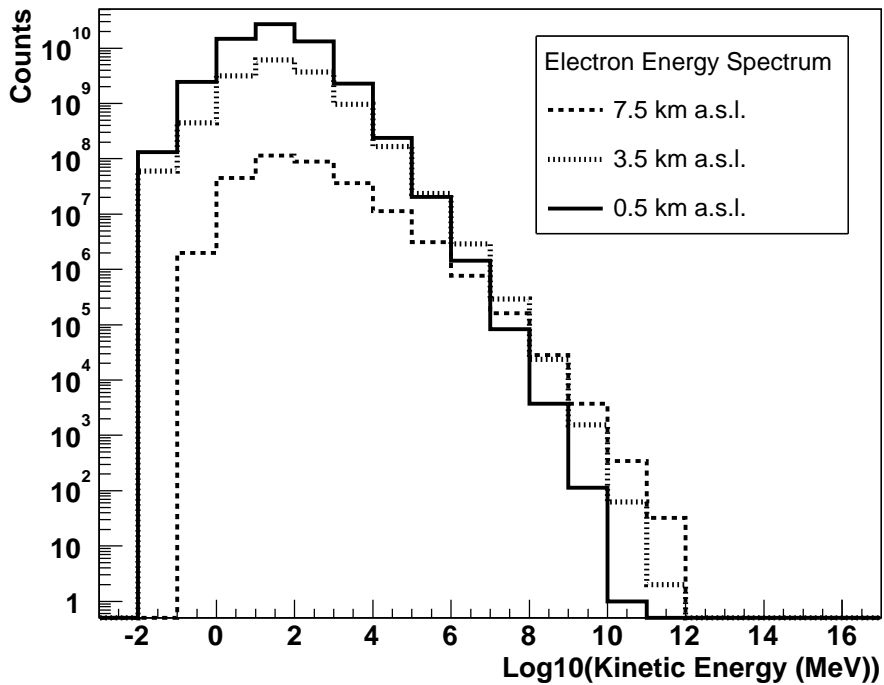}}
\caption{Examples of the electron kinetic energy distribution for ten
  proton vertical showers with energy  $10^{19}$ eV at
  0.5, 3.5 and 7.5 km above sea level.}  
\label{fig:spec:elec}
\end{figure}

Figure \ref{fig:spec:elec} shows an example of the kinetic energy
distribution electrons of ten  showers initiated by proton with
energy $10^{19}$ eV at 0.5, 3.5 and 7.5 km above sea level. It is
clear in this figure that the energy spectrum changes significantly
with height and that this must affect the dependence of the yield with
height.

In order to define a characteristic energy for which we could get a
fluorescence yield that represents the energy spectrum of electrons we
should solve the following equation:

\begin{equation}
\int S(K) \times FY(K,\rho,T)) \  dK =   FY(K_{ch},\rho,T) \times \int S(K) \  dK
\label{eq:yield:solve}
\end{equation}

where in the left hand side we have the energy spectrum ($S(K)$) convoluted
with the fluorescence yield ($FY(K,\rho,T)$) and in the right hand side we have
the yield calculated at a characteristic kinetic energy ($K_{ch}$)
times an average energy. Using the proportionality of the fluorescence
yield with the $dE/dx$ we can solve the equation and find the values
of the characteristic energy which we can use in order to calculate a
representative fluorescence yield.

\begin{figure}[]
\centerline{\includegraphics[width=13cm]{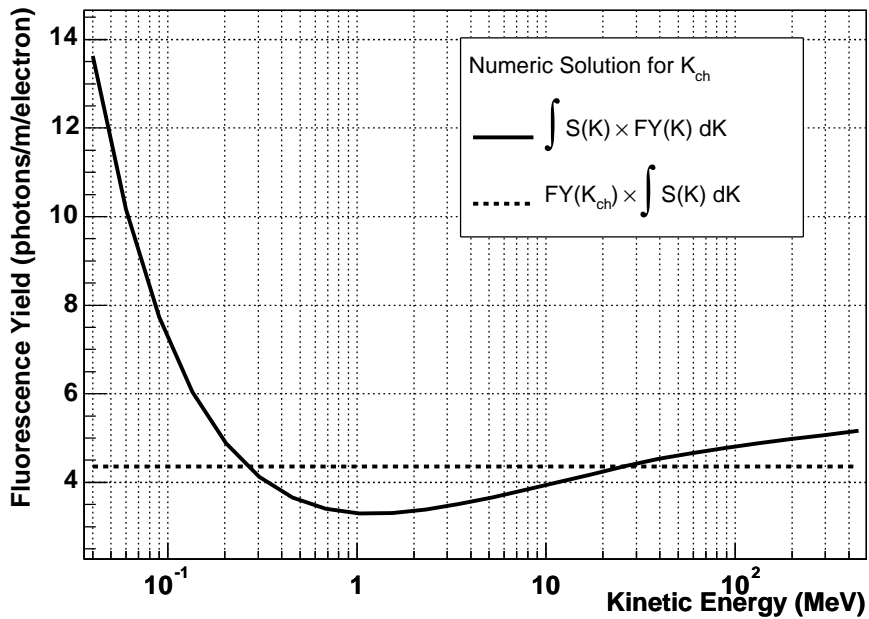}}
\caption{Numeric solution of equation \ref{eq:yield:solve} for the
  electron spectrum for showers initiated by protons with $10^{19}$ 
  eV at 1.5 km  a.s.l..} 
\label{fig:yield:solve}
\end{figure}

Figure \ref{fig:yield:solve} shows an example of the numeric solution
for equation \ref{eq:yield:solve} for the energy spectrum of
electrons in proton showers at 1.5 km a.s.l. (above sea level)
supposing the correspondence of the yield with the energy deposit suggested by
Kakimoto et. al.

Equation \ref{eq:yield:solve} has two solutions with the same
representative yield. We are going to quote always the higher kinetic
energy. 
In case of figure \ref{fig:yield:solve} the intersection occurs at
$K_{ch} = 25.8$ MeV what leads to a representative yield of 4.36
photons/m/electron.

The same calculation was repeated for all heights, different primary
particles and energies.

\subsection{Varying primary particles type and energy}

For different primary particle types and energies the electron energy
 spectrum  at a given height may be different and therefore the
characteristic energy could also change. In order to check the later, we have
simulated sets of ten showers initiated by proton and iron nuclei
with primary energies of $10^{18}$, $10^{19}$ and $10^{20}$ eV.

\begin{figure}[]
\centerline{\includegraphics[width=13cm]{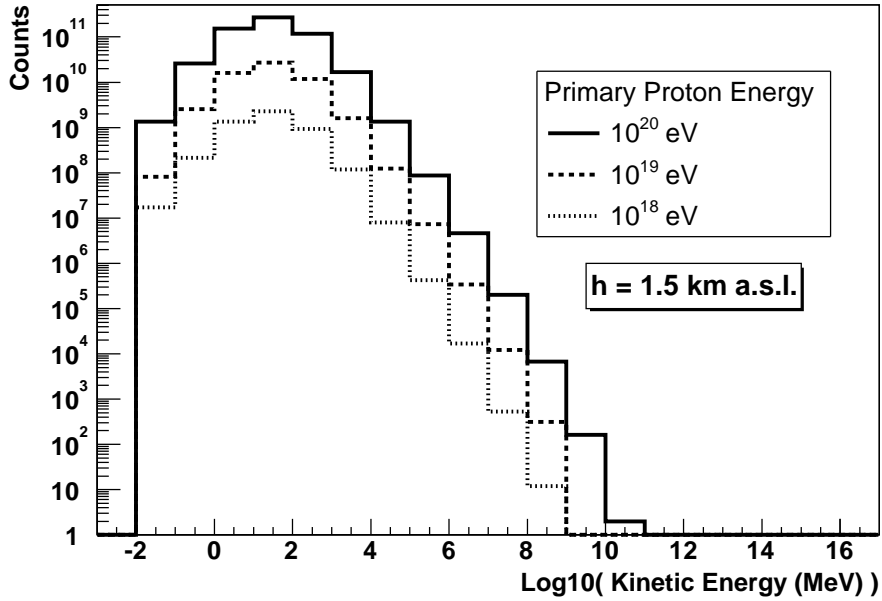}}
\caption{Kinetic energy spectrum for electrons at 1.5 km in ten
  vertical showers initiated by proton with energies $10^{18}$,
  $10^{19}$ and $10^{20}$ eV.}  
\label{fig:espec:ener:pr}
\end{figure}

Figure \ref{fig:espec:ener:pr} shows the electron energy spectrum 
for proton showers at different primary energies at 1.5 km
a.s.l.. This figure shows that the number of electron at a given
energy interval increases with energy independently of the kinetic
energy. In other words, the overall shape of the spectrum is the same for
different primary energies what directly implies that the fluorescence
yield is not a strong function of the energy for primary
protons. Exactly the  same feature happens for primary iron showers
where the fluorescence yield does not change very much with energy.

\begin{figure}[]
\centerline{\includegraphics[width=13cm]{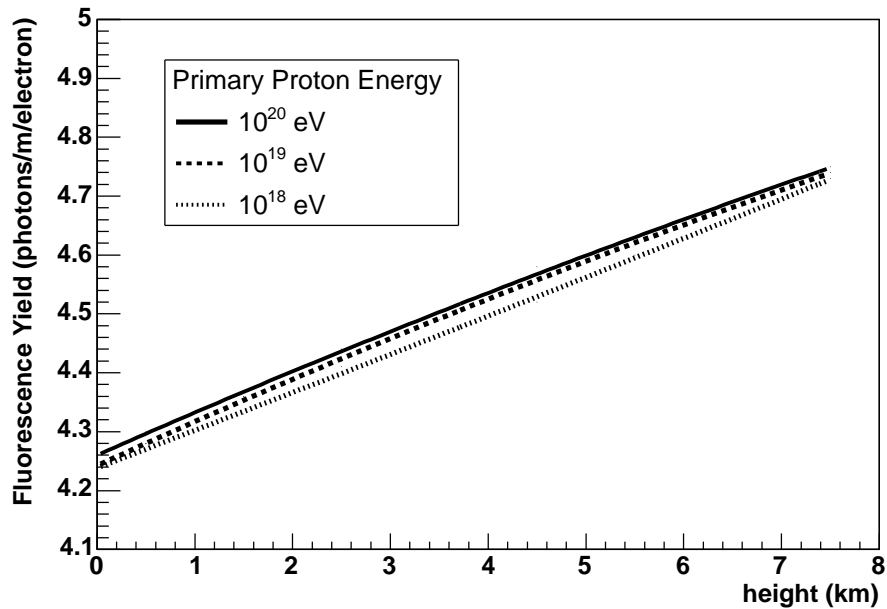}}
\caption{Fluorescence yield according to Kakimoto et al. suggestion as
  a function of height calculated taking 
  into account the energy spectrum as explained above for  vertical
  showers initiated by proton with energies $10^{18}$, 
  $10^{19}$ and $10^{20}$ eV.}  
\label{fig:yield:hei:en:pr}
\end{figure}

Despite that, small changes in the characteristic energy are seen for
different primary energies. For example, at 4.5 km for primary proton 
the characteristic energy is 28.6, 31.7 and 32.5 MeV for primary
energies of  $10^{18}$, $10^{19}$ and 
$10^{20}$ eV, respectively. However, since the $dE/dx$ curve vary
slowly with energy at this energy range, the correspondent yield changes are
within 0.1 photons/m/electron. The $dE/dx$ values are  4.52, 4.56 and
4.58 respectively.  Figure \ref{fig:yield:hei:en:pr} shows how the
fluorescence yield vary with height for proton primary showers with
different energies taking into account the energy spectrum as
explained above for the Kakimoto et al. correspondence between yield
and $dE/dx$. We remember that the Nagano et al. correspondence between
yield and $dE/dx$ is different only a constant factor.

On average we notice a tendency of the fluorescence yield to get
higher with increasing primary energy but we emphasize that the
differences are always smaller than 0.1 photons/m/electron. 

\begin{figure}[]
\centerline{\includegraphics[width=13cm]{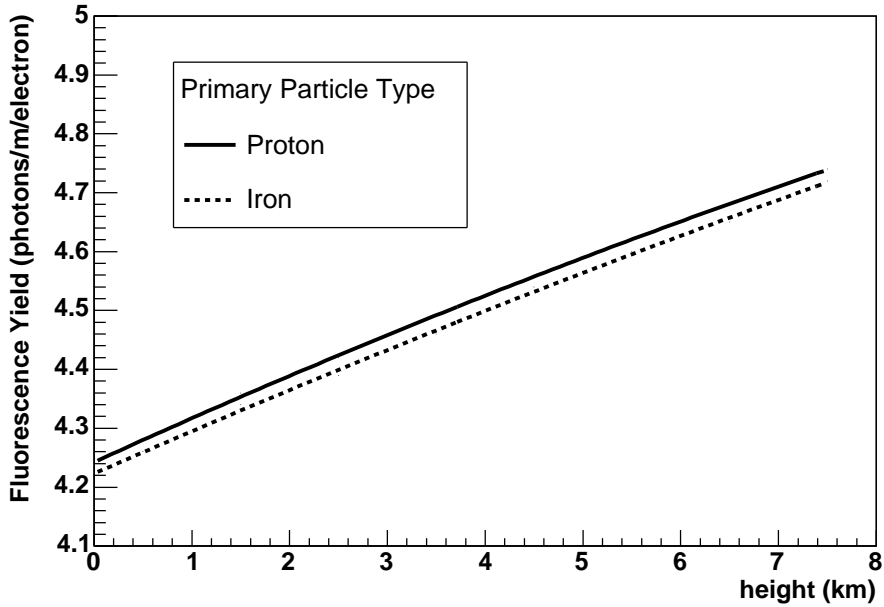}}
\caption{Fluorescence yield as a function of height calculated taking
  into account the energy spectrum as explained in section
  \ref{sec:analitic} according to Kakimoto et al. suggestion  for vertical
  showers initiated by proton and iron with energy $10^{19}$ eV.}  
\label{fig:yield:type}
\end{figure}

For iron showers we get the same results and conclusion except that
the differences between the fluorescence yield for the primary
energies are still smaller than the ones for proton showers. Figure
\ref{fig:yield:type} shows the small differences for the yield
calculated for primaries protons and iron nuclei with energy $10^{19}$
eV.

\subsection{The fluorescence yield for inclined showers}
\label{sec:inc}

The development of a shower is proportional to the amount of matter 
traversed in the atmosphere. Therefore, the energy spectrum of the
electrons in a vertical shower at 1.5 km is not equal to the electrons energy
spectrum in an inclined shower at 1.5 km. The average
electrons energy spectrum in 
showers with different inclinations is only equal at the same slant depth.

\begin{figure}[]
\centerline{\includegraphics[width=13cm]{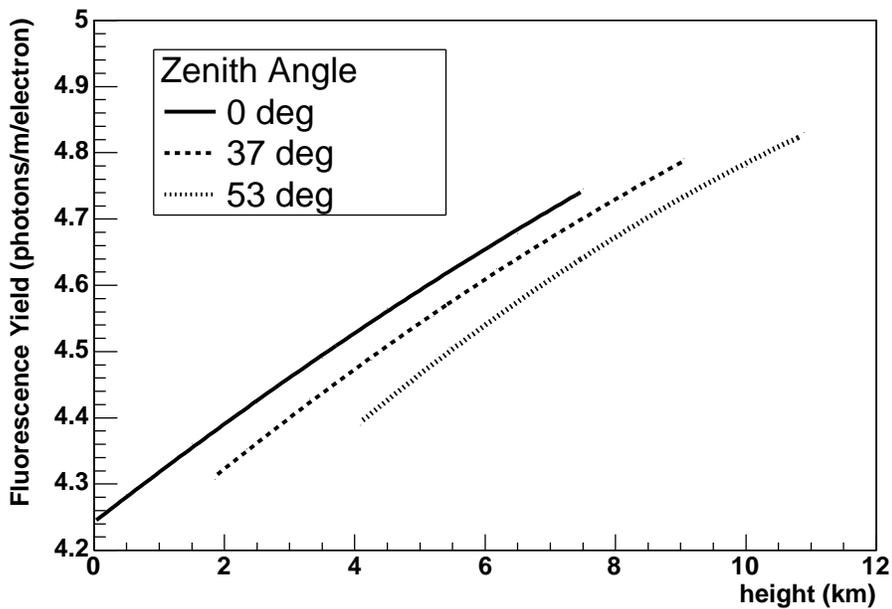}}
\caption{Fluorescence yield as a function of height calculated taking
  into account the energy spectrum as explained in section
  \ref{sec:inc} according to Kakimoto et al. suggestion for showers
  initiated by proton with energy $10^{19}$ 
  eV and zenith angle of 0, 37 and 53 degrees.}  
\label{fig:yield:inc}
\end{figure}

Figure \ref{fig:yield:inc} shows the fluorescence yield as a function
of height for proton showers with energy $10^{19}$ eV for three
inclinations.

The calculation regarding the energy spectrum has followed the same
procedure given in section \ref{sec:analitic}. To reproduce the
inclination of the shower we took the slant depth in the
vertical shower and transformed it to equivalent height in the inclined
shower. This is equivalent to suppose the shower has the same energy
spectrum at a given slant depth, what would lead to the same
characteristic energy. 

The characteristic energy, as can be seen in equation
\ref{eq:yield:solve}, is independent of pressure and temperature and
can be  given as a function of slant depth. However, the fluorescence
yield calculated for a given characteristic energy is a function of
pressure and temperature and hence must be calculated at a given
height for each shower inclination.

Figure \ref{fig:yield:inc} shows that for the same height the
fluorescence yield changes as much as 0.1 photons/m/electron both from
0 to 37 and from 37 to 53 degrees.

\subsection{Evaluating the fluorescence yield dispersion}

Another important feature of the fluorescence yield of electrons in an
air shower is their distribution. Historically, experiments have used an
average of the yield of all electrons at a given height
without considering the spread of the yield around this average value.

\begin{figure}[]
\centerline{\includegraphics[width=13cm]{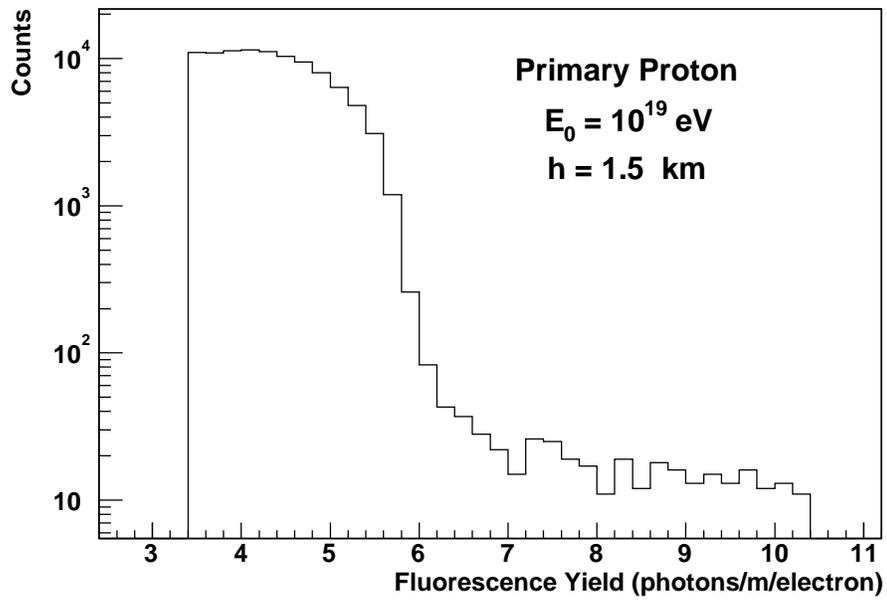}}
\caption{Distribution of the fluorescence yield calculated taking
  into account the energy spectrum as explained above according to
  Kakimoto et al. suggestion for vertical
  showers initiated by proton with energy $10^{19}$ eV.}  
\label{fig:yield:dist}
\end{figure}

At each height, we can drawn single electron energies following the
energy distribution probability as shown in  figure \ref{fig:spec:elec}.
We have drawn $10^5$ electron energies and for each one we calculated
the correspondent yield. Figure \ref{fig:yield:dist} shows an example of the distribution of
the yield for a proton shower with energy $10^{19}$ eV following
Kakimoto et al. suggestion. 

As expected, the distribution
has a pronounced peak but also has a long tail with high
yield values calculated for low energy electrons which is normally
neglected. There is an artificial maximum value for the fluorescence
yield which
corresponds to the minimum electron energy in the simulation (50 keV).

\section{Final Results and Conclusion}

In the reconstruction of cosmic ray showers with the fluorescence
technique it is impossible to know the energy and the identity of the
primary particle
before the application of the fluorescence yield. Therefore, the
safest procedure would be to use the average fluorescence yield
corresponding to the energies and primary particles we expect to
measure.

\begin{figure}[]
\centerline{\includegraphics[width=13cm]{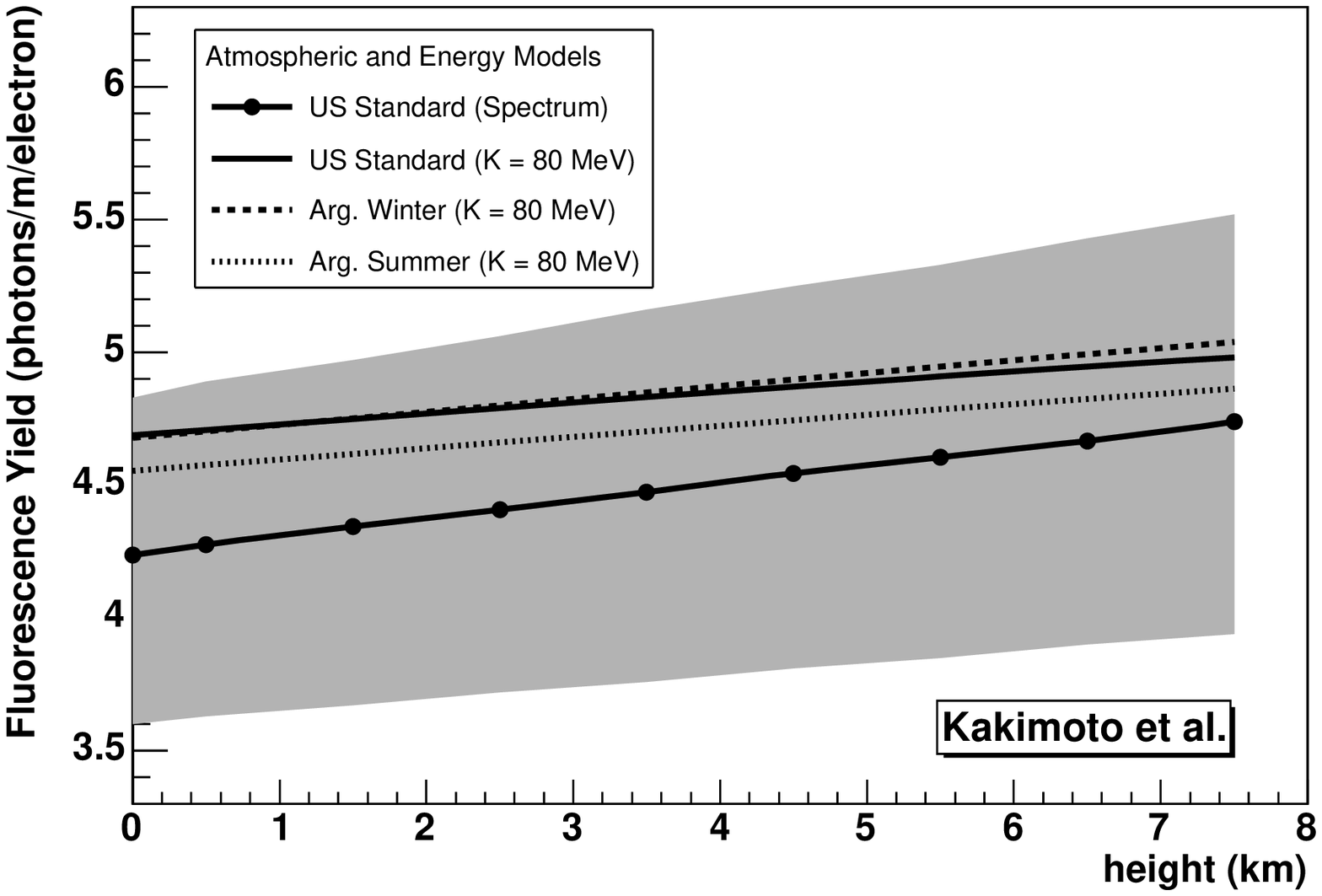}}
\caption{Average fluorescence yield as a function of height. The
  average was performed over 30 proton and 30 iron showers with
  energies $10^{18}$,  $10^{19}$ and $10^{20}$ eV.  It is
  shown the yield calculated according to the Kakimoto et al. equation
  for the US Standard, Argentine Winter and Argentine Summer for 80
  MeV electron and for comparison it is also shown the same
  calculation for the US Standard taking into account the electron
  energy spectrum. Hatched region corresponds to 68\% C.L..}
\label{fig:yield:height:kak}
\end{figure}

\begin{figure}[]
\centerline{\includegraphics[width=13cm]{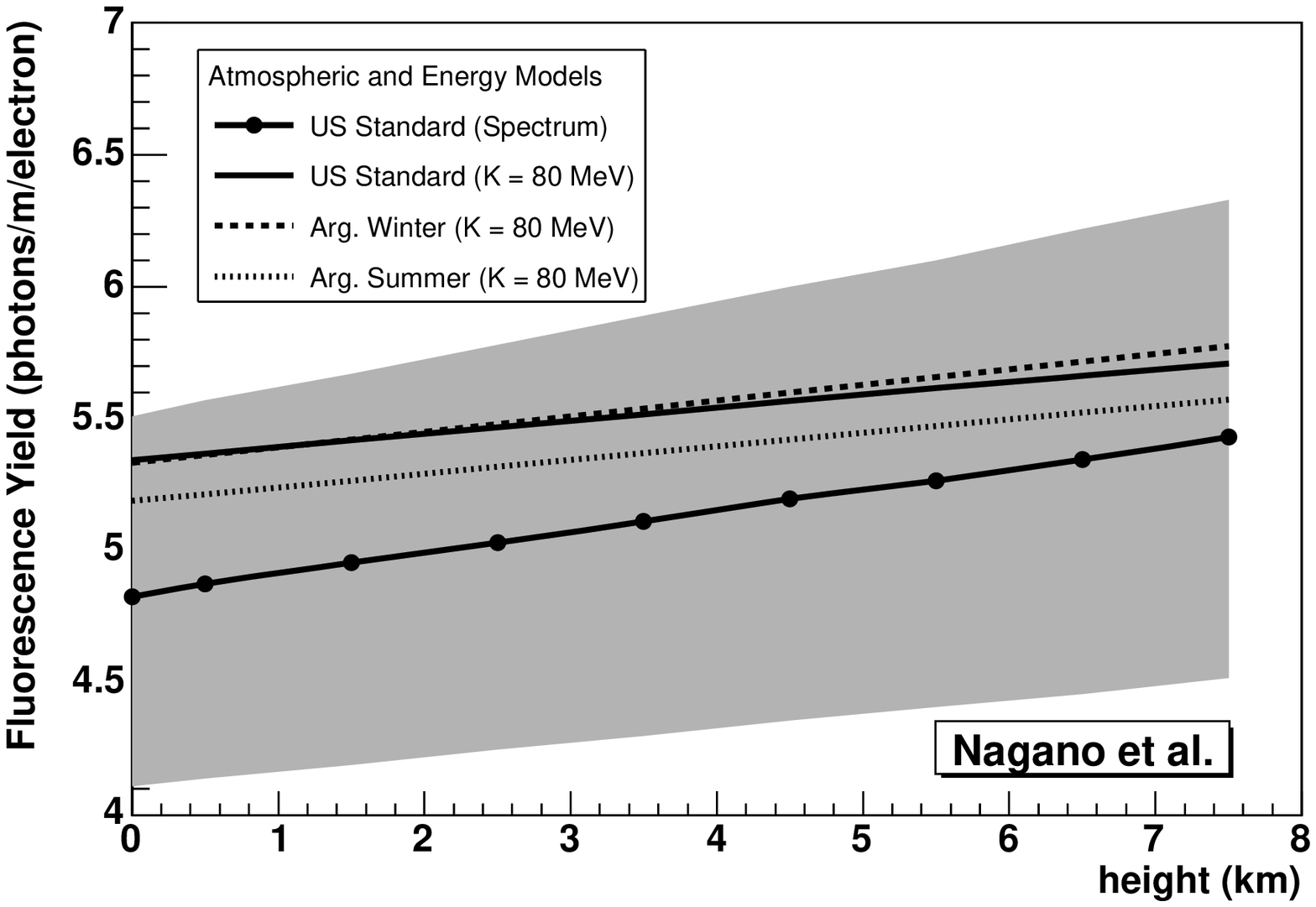}}
\caption{Average fluorescence yield as a function of height. The
  average was performed over 30 proton and 30 iron showers with
  energies $10^{18}$,  $10^{19}$ and $10^{20}$ eV. It is
  shown the yield calculated according to the Nagano et al. equation
  for the US Standard, Argentine Winter and Argentine Summer for 80
  MeV electron and for comparison it is also shown the same
  calculation for the US Standard taking into account the electron
  energy spectrum. Hatched region corresponds to 68\% C.L..}
\label{fig:yield:height:nag}
\end{figure}

Figures \ref{fig:yield:height:kak} and  \ref{fig:yield:height:nag}
show the average fluorescence yield as a function of height taking
into account the electron energy spectrum. We have averaged
over the 60 showers initiated by proton and iron with primary
energies of $10^{18}$, $10^{19}$ and $10^{20}$ eV. The hatched region shows
the 68\% confidence level of the distribution of the yields for each
height.

We also show in the same plot the fluorescence yield calculated for a
mono energetic electron (80 MeV) for three different atmospheres: US
Standard, Argentine Winter and Argentine Summer as measured by the
Auger Collaboration and published at \cite{bib:bianca:artigo}.

\begin{figure}[]
\centerline{\includegraphics[width=13cm]{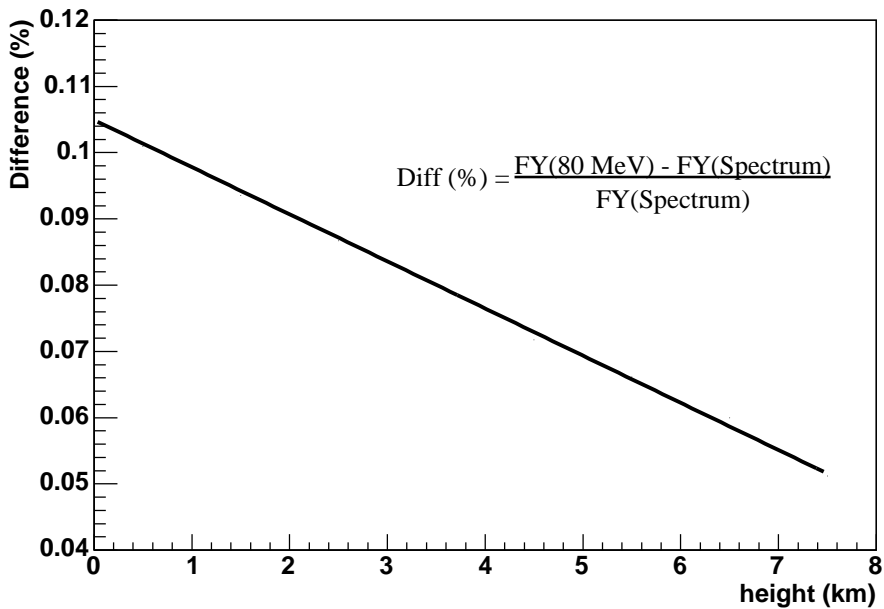}}
\caption{Difference between the fluorescence yield
  calculated according to the energy spectrum and calculated for a
  fixed energy (80 MeV) as a function of height.}
\label{fig:yield:diff}
\end{figure}

Only the US Standard atmosphere was used for these calculations. The
difference from the 
fixed energy to the spectrum calculations varies with height as shown in figure
\ref{fig:yield:diff} and the average difference  is
about 8\%. 

The exact value
the energy reconstructed using the mono energetic model (80 MeV) has
been overestimated varies from shower to shower but our results
suggests that fluorescence experiments should reevaluate their energy
estimative reducing them by $\sim$ 8\%. 

\begin{figure}[]
\centerline{\includegraphics[width=13cm]{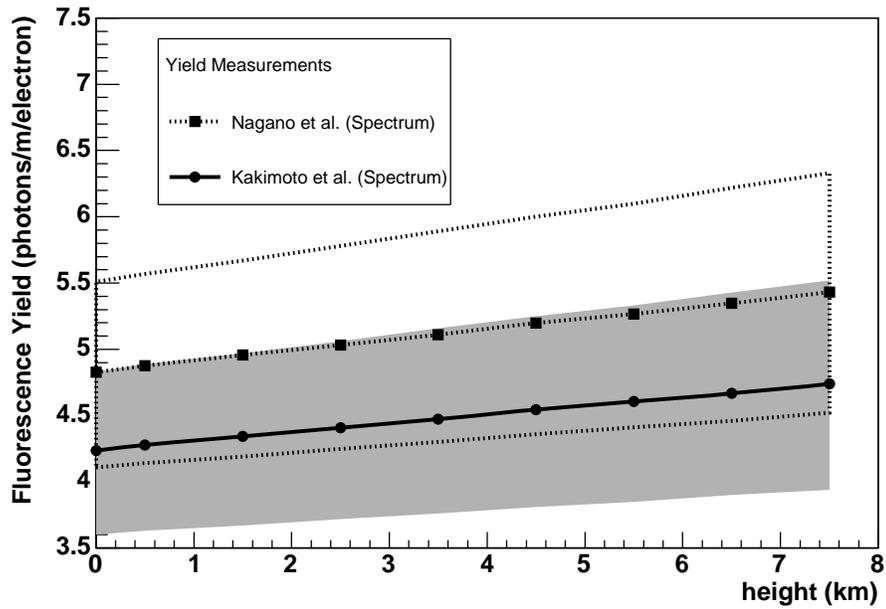}}
\caption{Average fluorescence yield as a function of height. It is
  shown the yield calculated according to the Nagano et al. and
  Kakimoto et al. measurements. We have used the
  US Standard atmosphere and the calculation was done
  taking into account the electron energy spectrum.}
\label{fig:final}
\end{figure}

The shower-to-shower and particle-to-particle fluctuation of the yield
due to the proper consideration of the kinetic 
energy distribution function of the
electrons is larger than 10\% which at least the
double of the variation due to atmospheric changes. The distribution
of the yield due to the energy distribution is so big that it would
 make the measurements of Kakimoto
et al. and Nagano et al. agree at the  68\% C.L. as can be seen in figure
\ref{fig:final}.

\begin{figure}[]
\centerline{\includegraphics[width=13cm]{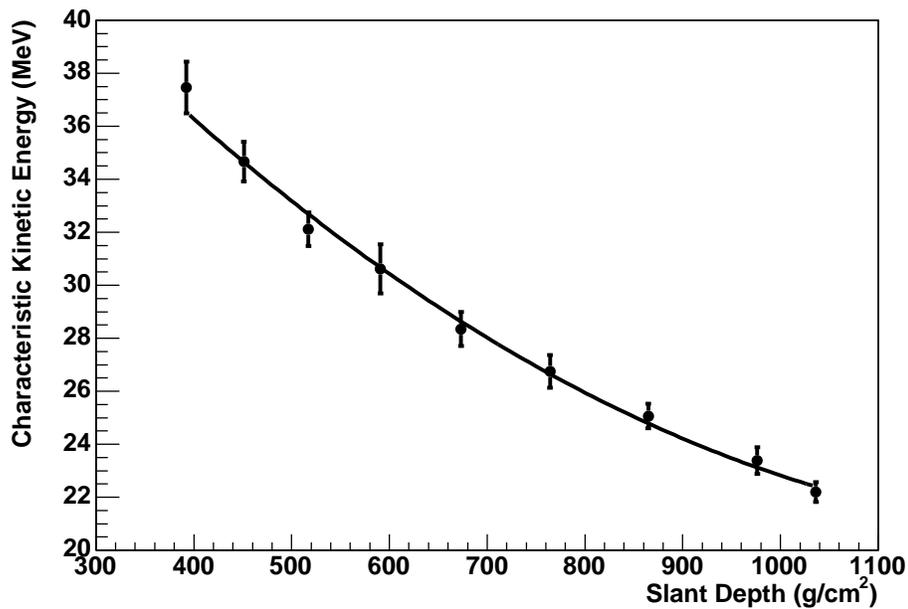}}
\caption{Average characteristic kinetic energy as a function of slant
  depth.  The
  characteristic energy is the average over 60 showers initiated by
  proton and iron nuclei with energies $10^{18}$, $10^{19}$ and $10^{20}$ eV.}
\label{fig:kch}
\end{figure}

The fluorescence yield to be applied to a cosmic ray shower depends on
the inclination of the axis of the shower. We define in equation
\ref{eq:yield:solve} the concept of characteristic energy which is the
kinetic energy for which the yield can be calculate to take into
account the electron energy spectrum in the shower. The
average characteristic energy for showers initiated by protons and
irons with energy  $10^{18}$, $10^{19}$ and $10^{20}$ eV is shown in
table \ref{tab:kch}. The characteristic energy is the natural value to
be used in the reconstruction of the shower as a function of slant
depth as it is also shown in figure \ref{fig:kch}.

Given the geometry of the axis of the shower,
at a given slant depth the characteristic energy should be used to
calculated the fluorescence yield as a function of height. Figure
\ref{fig:kch} shows the points calculated for each height and a fit to
these points. According to this simple fit the characteristic energy
can be given as a function of slant depth ($\chi$) by the equation:

\begin{equation}
K_{ch}(\chi) =  52.028\ - \ 0.04617 \  \chi \ + \ 1.6968\times 10^{-5}  \ \chi^{2}
\end{equation} 

We therefore conclude that the electron kinetic energy spectrum 
can not be neglected in order to
calculate the proper fluorescence yield to be used for energy
reconstruction of cosmic ray showers otherwise an over estimation of
8\% is artificialy introduced. 

\section{Acknowledgments}

This work was supported by the Brazilian population via the science
foundations FAPESP and CNPq to which we are grateful. 
V. de Souza is supported by FAPESP and J.A.Ortiz is supported by CNPq
by Post-Doc Fellowships.

\bibliographystyle{elsart-num}

\bibliography{yield}

\newpage 

\begin{table}[t]
\begin{center}
\begin{tabular}{|c|c|c|} \hline
Constants  &  Kakimoto et al. & Nagano et al. \\ \hline
$A_1 \ \mathrm{(m^2 \ kg^{-1})}$           & 89.0 $\pm$ 1.7  & 147.4 $\pm$ 4.3
\\ \hline    
$A_2 \ \mathrm{(m^2 \ kg^{-1})}$           & 55.0 $\pm$ 2.2  & 69.8 $\pm$ 12.2 \\ \hline    
$B_1 \ \mathrm{(m^3 \ kg^{-1} \ K^{-0.5}})$& 1.85 $\pm$ 0.04 & 2.40 $\pm$ 0.18 \\ \hline    
$B_2 \ \mathrm{(m^3 \ kg^{-1} \ K^{-0.5})}$ & 6.50 $\pm$ 0.33 & 20.1 $\pm$ 6.9 \\ \hline    
$K_c$  (MeV)                     & 1.4             & 0.85 \\ \hline    
\end{tabular}
\caption{Constants used by Kakimoto et al. and Nagano et al. in
  equation \ref{eq:yield}.}
\label{tab:const}
\end{center}
\end{table}

\begin{table}[t]
\begin{center}
\begin{tabular}{|c|c|} \hline
Slant Depth $\mathrm{(g/cm^2)}$ &  $K_{ch}$ (MeV) \\ \hline 
1036 &  22.2 $\pm$ 0.3 \\ \hline
976 & 23.4 $\pm$  0.5\\ \hline
864 & 25.1 $\pm$  0.5  \\ \hline
764 & 26.7 $\pm$  0.6 \\ \hline
673 & 28.3 $\pm$  0.6 \\ \hline
590 & 30.6 $\pm$  0.9 \\ \hline
517 & 32.1 $\pm$  0.6 \\ \hline
451 & 34.7 $\pm$  0.7 \\ \hline
392 & 37.5 $\pm$  0.9 \\ \hline
\end{tabular}
\caption{Characterist energy as a function of slant depth. The
  characteristic energy is the average over 60 showers initiated by
  proton and iron nuclei with energies $10^{18}$, $10^{19}$ and $10^{20}$ eV.}
\label{tab:kch}
\end{center}
\end{table}

\end{document}